# Spoof Surface Waves on Non Conducting Structured Interfaces

Mathieu Poulin, and Maksim Skorobogatiy, *Senior Member, IEEE*

*Abstract*— Spoof surface waves are demonstrated theoretically to propagate along periodic corrugated surfaces made of non-conductive lossless materials with positive permittivity. An analytic derivation of the Spoof surface wave dispersion relation is presented in the case of deeply subwavelength period of a corrugated structure using impedance boundary conditions at the interfaces. Thus obtained dispersion relation is verified numerically and limitations as well as physicality of impedance boundary conditions when modeling Spoof surface waves are discussed. Finally, suitable materials, potential experimental realization, and sensing applications of Spoof surface waves in the terahertz spectral range are discussed.

## I. Introduction

THE strong spatial confinement of surface waves near interfaces allows to greatly enhance performance of many practical devices in high-resolution sensing, spectroscopy, and subwavelength imaging [1-8]. Particularly, surface plasmon-polaritons propagation at the interface between a metal and an analyte revolutionized the field of bio-sensing in the visible spectral range [9-11]. Recently, THz (terahertz) spectral range emerged as the next frontier for non-destructive imaging and biomedical applications [12-19]. Because of the relative transparency of dry dielectrics to THz radiation and existence of spectral fingerprints for many biological molecules, THz sensing became an emerging research field that could also benefit from surface wave sensing modality. However in the long wavelength spectral range, excitation and application of surface plasmon-polariton waves has been problematic due to the THz wave weak confinement (strong presence in the cladding) and high losses [20, 21]. This is because in the THz spectral range imaginary part of the metal dielectric constant dominates and it can be several orders of magnitude larger than the real part unlike in the visible and near-IR spectral ranges, where real part of the metal dielectric constant is negative and much larger (in its absolute value) than the imaginary part. This entailed a search for new polaritonic materials capable of supporting THz surface waves with strong localisation at the interface [22-29]. One interesting alternative to surface plasmon-polaritons at long wavelengths are surface phonon-polaritons, which are excited on surfaces of some polar materials, typically in the near vicinity of material optical resonances. Such surface waves is a result of coupling between phonon and electromagnetic fields, with enabling materials abundant in the mid-IR and THz spectral ranges [30, 31] in the form of semiconductors and ceramics. Near optical material resonances, optical properties of surface phonon-polaritons in the THz spectral range can be similar to those of a surface plasmon polariton in the visible. This is because dielectric permittivity of a polaritonic material can feature a negative real dielectric constant that is much larger (in absolute value) than its imaginary part, which is similar to the case of metals in the visible and near-IR.

Recently, materials with subwavelength structuring were shown to offer an alternative to the traditional homogeneous materials for the excitation of surface waves as their interfacial optical properties are highly designable [32-34]. Particularly, structured metallic surfaces featuring periodic sequence of narrow air-filled grooves of various geometries such as rectangular [35-38], V-shape [39, 40], and radial [41] along with corrugated metallic wires [42-45] and helically grooved wires [46] were shown to support surface waves called Spoof plasmon-polaritons that feature strong field localization in the vicinity of metal even at long wavelengths. Additionally, corrugated surface made of semiconductors can also be designed to support Spoof surface waves [47, 48]. Remarkably, Spoof surface waves can be also excited on ideal lossless metallic surfaces characterized by the infinite negative dielectric constant (Perfect Electric Conductor) [49-52]. In practice, however, due to high material losses of metals in the THz spectral range, and due to tight modal confinement in the narrow metallic grooves, propagation losses of Spoof plasmon-polaritons tends to be large. One way to reduce the loss of a Spoof wave is to push its fields out into the low-loss air cladding by either using shallow grooves [53], metallic wedges [54] or by replacing a corrugated plate with a corrugated metallic stipe [55-59], in both cases, however, surface wave confinement near corrugated interface is decreased.

More generally, surface waves that are guided by conductive surfaces are known as Zenneck waves [60], and in this respect Spoof plasmon-polaritons propagating along surfaces of real lossy metals can be also called Spoof Zenneck waves. This nomenclature become even more relevant in the THz spectral range where metal dielectric constant is dominated by the large

Manuscript received MONTH DATE, 2021; revised MONTH DATE, 2021; accepted MONTH DATE, 2021. Date of publication MONTH DATE, 2021; date of current version MONTH DATE, 2021. We would like to acknowledge financial support of the Canada Research Chair Tier I program of Prof. Skorobogatiy in the area of Ubiquitous Terahertz Photonics. The authors are with the Department of Engineering Physics, Polytechnique de Montreal, Montréal, QC H3C 3A7, Canada (e-mail: maksim.skorobogatiy@polymtl.ca; mathieu-2.poulin@polymtl.ca).

imaginary part. More importantly, a detailed analysis shows that for a planar surface to support a guided wave, it is only necessary for the underlying material to be lossy (have a non zero imaginary part of the dielectric constant), while the sign of a real part of the dielectric constant is irrelevant [21, 61]. One, therefore, further distinguishes surface waves propagating on lossy materials having positive real part of the dielectric constant as Brewster waves. Over the years, there was much discussion about whether Zenneck and Brewster surface waves are well defined and relevant in practical applications. The problem stems from the realisation that reliable experimental characterization of such waves normally requires their relatively strong localization at the interfaces (transverse modal sizes of several tens to hundreds of wavelengths at most) to prevent significant interaction with the environment. Moreover, it is also desirable for the surface wave propagation distances to be much longer than the wave transverse size, as in this case surface wave resembles a guided mode that can be excited and probed much like a regular guided mode of a waveguide. Unfortunately, for many real materials and practical operation conditions Zenneck and Brewster waves are either extremely extended into the cladding and thus easily lost to the environment, or they feature very high transmission losses that result in the surface wave propagation length comparable to the wave transverse size. For these reasons, such surface waves are often difficult to excite and classify as true guided modes rather than some kind of local geometrical resonances.

In this paper, we demonstrate for the first time to our knowledge that the well-defined Spoof surface waves can propagate on structured non-conductive surfaces made of lossless dielectrics featuring real positive dielectric permittivities. Particularly, we show that using common high permittivity materials and judicious choice of the surface geometry one can realise strong confinement of Spoof waves near the interface, as well as propagation distances that greatly exceed the wave transverse size. Moreover, under certain conditions, Spoof surface waves can also propagate along structured surfaces made of a material with infinite positive dielectric constant (Perfect Isolator). In this respect, the surface waves described in our work can also be classified as Spoof Zenneck or Spoof Brewster waves.

More precisely, we study corrugated surfaces featuring 1D periodic arrays of grooves filled with lossless low-permittivity dielectric of the upper cladding. Such grooves are separated by notches made of lossless high-permittivity dielectric of the lower cladding. Dispersion relation of Spoof surface waves are obtained semi-analytically in the long wavelength limit by solving Maxwell equations in the vicinity of a structured surface following a general approach detailed in [36, 62]. One key difference of our derivation compared to the ones presented in the abovementioned refences is that we use Impedance Boundary Conditions (IBC) at the interface with a structured interface, as well as positive values of the material dielectric constants [63]. This also entails an important and unique for this problem discussion about the effect of radiation leakage from the grooves into the notches, potential dissipation mechanisms of such leaked radiation inside the notches, as well as necessity for suppression of radiation-induced coupling between adjacent grooves. Semi-analytical results with IBC were then confirmed with numerical simulations using COMSOL Multiphysics Software with IBC, and very good correspondence was observed. Finally, a full simulation using COMSOL using two cladding materials and a structured interface confirmed that our semi-analytical theory indeed describes well Spoof surface waves on non-conductive surfaces with the exception of resonant frequencies at which structural resonances are excited within the high dielectric permittivity notches. In principle, such resonances can be removed by various techniques as described further in the paper. Although our derivations are not wavelength specific, without the loss of generality we present examples in the THz spectral range (0.1 to 1 THz) in view of our future experimental work on optical characterization of such waves.

## II. DERIVATION OF THE DISPERSION RELATION

Consider a periodically corrugated surface made of a material with real and positive permittivity $\varepsilon_m > 0$ adjacent to a lossless dielectric filler of permittivity $\varepsilon_0 > 0$ as shown in Fig. 1. This periodic structure has a period $d$ and contains grooves of width $a$ and height $h$. In what follows we call notches the rectangular regions of length $d - a$ between grooves that are filled with material $\varepsilon_m$. Both materials are assumed to be non-magnetic. The dispersion relation of a Spoof surface wave supported by the interface can be found by solving the Maxwell equations in the semi-infinite upper zone II ($z > 0$) and grooves in the lower zone I ($z < 0$), while respecting the continuity conditions of the fields at $z = 0$. Additionally, the impedance boundary condition (IBC) is used at the interfaces with the high dielectric material (zone III) in the following form [64]

$$\eta_s \boldsymbol{s} \times \boldsymbol{H} + \boldsymbol{s} \times (\boldsymbol{s} \times \boldsymbol{E}) = 0, \qquad (1)$$

where $\eta_s$ is the surface impedance and $\boldsymbol{s}$ is the unit normal vector to the interface. This simplest form of IBC can be derived by considering a planewave incident from the material with lower dielectric permittivity onto a planar interface with material with higher dielectric permittivity. The incident E&M fields satisfy (1) at the interface, assuming that only an outgoing wave is present on the side of a higher permittivity dielectric. This boundary condition is attractive in its simplicity, but if it is to be of universal nature and practical utility, it is necessary that the surface impedance $\eta_s$ be independent of the incident field characteristics, such as polarization and angle of the incident plane wave. Following [20] one can demonstrate that if the two materials occupying the half spaces are homogeneous and isotropic, and if the incident wave comes from material $\varepsilon_0$, while for the second material $|\varepsilon_m| \gg \varepsilon_0$, then surface impedance has the following simple form $\eta_s = 1/\sqrt{\varepsilon_m}$ (CGI units) for any polarization and incidence angle of the incoming wave. In the case of finite-size structures, such as sequence of grooves shown in Fig. 1 (a), in order for the impedance boundary condition to be valid at the interfaces with a filler dielectric, one has to demand that inside the notches only waves outgoing from the interfaces (into the notches) are excited, while no reflections from other surfaces reach those interfaces. Physically, this requires that waves propagating within the notches are either absorbed within the notch length, or that they are efficiently scattered within a notch and into the semi-infinite zone III, or that their coherence is somehow destroyed over the propagation distance within the notches via structured absorption of destructive interference as illustrated in Fig. 1 (b)-(e). Note that when treating Spoof Plasmon waves on corrugated metallic surfaces, validity of IBC is easy to ascertain as long as the notch width is much larger that the metal skin depth. In this case, a wave transmitted into metal from the filler dielectric will decay exponentially fast from the interface into the notch and will have no presence at other surfaces. In contrast, in the case of Spoof surface waves propagating on corrugated surfaces of lossless dielectrics, for IBC to be valid, it is not sufficient to demand that $\varepsilon_m \gg \varepsilon_0$, in addition, one has to ensure that experimental system includes one of the energy dissipation mechanisms described earlier that would prevent interaction between different interfaces of the same notch.

We now present brief derivation of the dispersion relation of Spoof surface waves propagating on corrugated interfaces of lossless dielectrics. First, we start with zone I that defines a planar waveguide filled with low-permittivity $\varepsilon_0$ material, and that supports modes propagating along OZ direction. The modes are confined between the walls of high-permittivity $\varepsilon_m$ material. One end of the waveguide is opened into zone II, while the other one is bound by a high-permittivity material. Assuming validity of the Impedance Boundary conditions, one can solve the Maxwell equations and find dispersion relation of the groove waveguide even modes described by the propagation constant $\beta_z^I$ that satisfies

$$k_x^I \tan(k_x^I a/2) = -i \frac{k_0 \varepsilon_0}{\sqrt{\varepsilon_m}}, \quad (2)$$

with $\beta_z^I = \pm\sqrt{\varepsilon_0 k_0^2 - (k_x^I)^2}$, where for a forward propagating wave (along OZ axis) $\text{Re}(\beta_z^I) > 0, \text{Im}(\beta_z^I) > 0$. Then, adapting an approach from the theory of Spoof Plasmons [65], in the limit of $dk_0\sqrt{\varepsilon_0} \ll 1$ (subwavelength corrugation), an analytic expression for the propagation constant $\beta_x^{SW}$ of a Spoof surface wave propagating on the corrugated interface along the OX axis can be found by solving Maxwell equations in the zone II, while matching the fields with those of zone I and respecting the IBC along $z = 0$ on the notch surfaces

$$\left(\frac{\varepsilon_0 k_0}{\sqrt{\varepsilon_m}} + k_z^{SW,II}\right)\frac{\sin\left(\beta_x^{SW}\frac{d}{2}\right)}{\sin\left(\beta_x^{SW}\frac{a}{2}\right)} = i\beta_z^I \frac{\sqrt{\varepsilon_m}\beta_z^I \tan(\beta_z^I h) + i\varepsilon_0 k_0}{\sqrt{\varepsilon_m}\beta_z^I - i\varepsilon_0 k_0 \tan(\beta_z^I h)} + \frac{\varepsilon_0 k_0}{\sqrt{\varepsilon_m}} \quad (3)$$

where $k_z^{SW,II} = \pm\sqrt{\varepsilon_0 k_0^2 - (\beta_x^{SW})^2}$ is the transverse (z-component) of the wave vector of a Spoof surface wave in the filler dielectric (zone II), while $k_z^{SW,III} = \pm\sqrt{\varepsilon_m k_0^2 - (\beta_x^{SW})^2}$ is the transverse (z-component) of the wave vector of a Spoof surface wave in the high permittivity dielectric (zone III) (see more details of derivation in the Supplemental Materials section). Although the derivations presented above assume a non-conductive material with purely real and positive value of $\varepsilon_m$, they also remain valid in case of lossy materials as long as $|\varepsilon_m| \gg \varepsilon_0$. Numerical methods such as the Newton-Raphson iteration method can be used to solve (2) and (3) together. In addition, in the most general case, the two dispersion relations (2)-(3) are universal functions of several non-dimensional variables namely $hk_0\sqrt{\varepsilon_0}$, $ak_0\sqrt{\varepsilon_0}$, $dk_0\sqrt{\varepsilon_0}$, and $\sqrt{\varepsilon_0}/\sqrt{\varepsilon_m}$. Indeed, defining Spoof surface wave effective refractive index as $n_{eff}^{SW} = \beta_x^{SW}/k_0$ and that of a groove waveguide as $n_{eff}^I = \beta_z^I/k_0$, as follows from the explicit forms of (2) and (3)

$$\frac{n_{eff}^I}{\sqrt{\varepsilon_o}} = f_{n_{eff}^I}\left(ak_0\sqrt{\varepsilon_0}, \frac{\sqrt{\varepsilon_0}}{\sqrt{\varepsilon_m}}\right) \quad (4)$$

$$\frac{n_{eff}^{SW}}{\sqrt{\varepsilon_o}} \simeq f_{n_{eff}^{SW}}\left(hk_0\sqrt{\varepsilon_o}, ak_0\sqrt{\varepsilon_0}, dk_0\sqrt{\varepsilon_o}, \frac{\sqrt{\varepsilon_0}}{\sqrt{\varepsilon_m}}\right). \quad (5)$$

We will use equations (4) and (5) later in the paper to evaluate the effects of geometrical and material parameters of the structured interface on the Spoof surface wave dispersion relation. We note that in addition to relations (2) and (3), dispersion relation of a Spoof surface wave must satisfy several physical conditions

$$\text{Re}(\beta_x^{SW}) > 0, \text{Im}(\beta_x^{SW}) > 0 \quad (6)$$

as all the materials in the system have real and positive values of the dielectric constants and no coherent back reflection is expected for corrugation with a deeply subwavelength period. At the same time, a properly defined physical solution requires that electromagnetic energy be localized near the corrugated surface while decaying away from the interface in both zone II and zone III. Thus, the signs of $k_z^{SW,II}$ and $k_z^{SW,III}$ in (2) and (3) must be chosen so that

$$\text{Im}(k_z^{SW,II}) > 0, \text{Im}(k_z^{SW,III}) < 0. \quad (7)$$

One generally finds that in case of a forward travelling Spoof surface wave with dispersion relation respecting (6) and (**7**), and similarly to the case of a classic Brewster wave [21], the wave phase front in zone II is traveling toward the interface, while in zone III it travels away from the interface

$$\text{Re}(k_z^{SW,II}) < 0, \text{Re}(k_z^{SW,III}) < 0. \quad (8)$$

This is also the reason why Brewster waves are closely related to the phenomenon of a Brewster angle [66].

Modal propagation distance $L_x$, as well as modal extent into the filler dielectric $L_z^{II}$ can be defined as distances over which the fields decay by a factor $e$ in the corresponding directions (+X, +Z or -Z)

$$L_x = 1/|\text{Im}(\beta_x^{SW})| \quad (9)$$

$$L_z^{II} = 1/|\text{Im}(k_z^{SW,II})|. \quad (10)$$

We note that for a Spoof surface wave to be properly defined one must require that its transverse sizes be much smaller than the propagation distance $L_z^{II} \ll L_x$.

### A. Case of infinite positive permittivity $\varepsilon_m = +\infty$

In this section, a case of a corrugated structure made of a perfect isolator ($\varepsilon_m = +\infty$) will be considered in order to find the Spoof surface wave cut-off frequencies and its dispersion in this simple limit. By considering $\varepsilon_m = +\infty$ one can greatly simplify the analytic dispersion relations (2) and (3)

$$\beta_z^I = \pm\sqrt{\varepsilon_0 k_0^2 - \left(\frac{2\pi}{a}l\right)^2}, \quad l = 0,1,2\ldots \quad (11)$$

$$k_z^{SW,II}\frac{\sin\left(\beta_x^{SW}\frac{d}{2}\right)}{\sin\left(\beta_x^{SW}\frac{a}{2}\right)} = i\beta_z^I \tan(\beta_z^I h). \quad (12)$$

These expressions allow to find frequency regions of Spoof surface wave existence. Thus, a new branch of the Spoof surface wave dispersion relation will be pulled from the radiation continuum (lower cut-off frequency) when $k_z^{SW,II} = 0 \rightarrow \beta_z^I h = \pi \cdot l_c$, while the same branch will be terminated at higher frequencies (upper cut-off frequency) when Spoof surface wave propagation constant becomes infinite when $\tan(\beta_z^I h) = \infty \rightarrow \beta_z^I h = \pi/2 + \pi \cdot l_c$. From (11) we then find the following expression for the frequency regions of existence of the Spoof surface waves

$$\nu_c \in \frac{c}{\sqrt{\varepsilon_0}}\left(\sqrt{\left(\frac{l}{a}\right)^2 + \left(\frac{l_c}{2h}\right)^2}, \sqrt{\left(\frac{l}{a}\right)^2 + \left(\frac{l_c+0.5}{2h}\right)^2}\right); \quad l = 0,1,2\ldots, \; l_c = 0,1,2\ldots, \quad (13)$$

while for the fundamental branch ($l = 0; l_c = 0$) it translates into $\nu_c \in \left(0, c/(4h\sqrt{\varepsilon_0})\right)$.

Expression (12) can be further simplified in the regime of subwavelength corrugation $|\beta_x^{SW}|d \ll 1$, where Spoof surface wave dispersion relation and its penetration depth into the filler dielectric $L_z^{II}$ can be written as

$$\beta_x^{SW} = \sqrt{\varepsilon_0 k_0^2 + \left[\beta_z^I \tan(\beta_z^I h)\left(\frac{a}{d}\right)\right]^2};$$
$$\varepsilon_{eff}^{SW} \simeq \varepsilon_0 + \left[\frac{a}{d}\frac{\beta_z^I}{k_0}\tan(\beta_z^I h)\right]^2 \quad (14)$$
$$L_z^{II} = \frac{1}{\beta_z^I \tan(\beta_z^I h)\left(\frac{a}{d}\right)}. \quad (15)$$

From these expressions it follows that the Spoof surface wave confinement in the filler dielectric can become subwavelength $L_z^{II} k_0 < 1$ if certain conditions on the groove depth are satisfied. Thus, for the fundamental mode of a groove $\beta_z^I = \sqrt{\varepsilon_0} k_0$, condition of subwavelength confinement at the surface vicinity in the filler dielectric becomes

$$\tan(\sqrt{\varepsilon_0} k_0 h) > \left(\frac{d}{a}\right)\frac{1}{\sqrt{\varepsilon_0}}, \quad (16)$$

signifying that subwavelength confinement is observed mostly in the vicinity of an upper cut-off frequency. In Fig. 2(a) (dashed curve) we present a typical dispersion relation of a Spoof surface wave using analytical dispersion relation (14) for $d = 1.5a$ while assuming that a subwavelength corrugation $|\beta_x^{SW}|d \ll 1$ holds for all the parameters and all frequencies. A subwavelength corrugation period $d = 45$ μm $< \lambda_0/5$ along with the depth of the grooves $h = 75$ μm are chosen so that the upper cut-off frequency of the fundamental branch is at $\nu_c = 1$ THz ($\lambda_0 = 300$ μm), while we suppose that filler dielectric is air $\varepsilon_0 = 1$. In solid curve we present numerical solution of (11), (12) solved with the Newton-Raphson iteration method for the same structure. The Spoof surface wave penetration depths $L_z^{II}/\lambda$ into the filler dielectric as calculated using the dispersion relation in the long wavelength limit (14) (solid curve), and numerically solved one (12) (dashed curve) are presented in Fig. 2(c). The subwavelength localization of the Spoof surface wave near the interface is possible near the upper cut-off frequency. Finally we note that when approaching the upper cut-off frequency (gray region in Fig. 2 (a)), solution of (11), (12) might too give high values of the Spoof surface wave propagation constant $|\beta_x^{SW}|d \gtrsim 1$, which eventually results in breaking down of the long wavelength approximation that was used to derive equations (3), (12). Thus, when tighter confinement of a Spoof wave near surface is desired, one has to resort to smaller periods $d$, that will allow operation closer to the upper cut-off frequency.

### B. Case of finite positive permittivity $\varepsilon_m$

Governing equations for the dispersion relation of a Spoof surface wave with finite permittivity $\varepsilon_m$ as given by (2) and (3) can be simplified in various limits. First of all, we note that in case of subwavelength corrugation $dk_0 \ll 1$, equation (2) can be solved analytically via linearization as $|k_x^I|a/2 \ll 1$ for any choice of $a$, and $k_0$ (indeed, linearization of (2) only requires that $ak_0 \ll 2\sqrt{\varepsilon_m}/\sqrt{\varepsilon_0}$). Furthermore, assuming a long wavelength limit $|\beta_x^{SW}|d \ll 1$ we get an analytic expression for the effective permittivity of a lossy fundamental Spoof surface wave

$$\beta_z^I = \sqrt{\varepsilon_0 k_0^2 + i\frac{2k_0\varepsilon_0}{a\sqrt{\varepsilon_m}}} \quad (17)$$

$$\varepsilon_{eff}^{SW} \simeq \varepsilon_0 - \left[\frac{a}{d}\frac{\beta_z^I}{k_0}\frac{\varepsilon_0 k_0 - i\sqrt{\varepsilon_m}\beta_z^I \tan(\beta_z^I h)}{\sqrt{\varepsilon_m}\beta_z^I - i\varepsilon_0 k_0 \tan(\beta_z^I h)} + \frac{\varepsilon_0}{\sqrt{\varepsilon_m}}\left(\frac{d-a}{d}\right)\right]^2. \quad (18)$$

Note that in this regime, the two dispersion relations are only functions of a few non-dimensional parameters namely $hk_0\sqrt{\varepsilon_o}$, $a/d$, $\sqrt{\varepsilon_0}/\sqrt{\varepsilon_m}$, and $1/ak_0\sqrt{\varepsilon_m}$ such as

$$\frac{n_{eff}^I}{\sqrt{\varepsilon_o}} = f_{n_{eff}^I}\left(\frac{1}{ak_0\sqrt{\varepsilon_m}}\right) \quad (19)$$

$$\frac{n_{eff}^{SW}}{\sqrt{\varepsilon_0}} \simeq f_{n_{eff}^{SZ}}\left(hk_0\sqrt{\varepsilon_o}, \frac{a}{d}, \frac{\sqrt{\varepsilon_0}}{\sqrt{\varepsilon_m}}, \frac{1}{ak_0\sqrt{\varepsilon_m}}\right) \quad (20)$$

Understanding validity condition $|\beta_x^{SW}|d \ll 1$ for equation (18) is somewhat complicated as in the vicinity of the upper cut-off frequency $\beta_z^I h \sim \pi/2$, the values of $\tan(\beta_z^I h)$ become large and the Spoof surface wave effective refractive index can also become large ($n_{eff}^{SW} \sim \sqrt{\varepsilon_m}$) both in its real and imaginary parts. Thus, if (18) is to be valid for any choice of parameters, $a, d, h$ in the vicinity of a cut-off frequency $\beta_z^I h \sim \pi/2$, then one has to require that $ak_0 \ll 1/\sqrt{\varepsilon_m}$. This requirement of extremely subwavelength grooves can lead to significant losses of Spoof surface waves due to inefficient confinement, and as a consequence, high radiation losses in such grooves as follows from (17). In practice, in order to minimize radiation loss in the grooves one would rather choose $ak_0 \gg 1/\sqrt{\varepsilon_m}$, therefore equation (18) is not expected to be valid in the immediate vicinity of the upper cut-off frequency $\beta_z^I h \sim \pi/2$, while it will be generally valid somewhat away from the cut-off frequency. Thus, assuming $ak_0 \gg 1/\sqrt{\varepsilon_m}$ and $\sqrt{\varepsilon_0}/\sqrt{\varepsilon_m} \ll \tan(hk_0\sqrt{\varepsilon_o}) \ll \sqrt{\varepsilon_m}/\sqrt{\varepsilon_0}$, equations (17) and (18) can be further simplified to give

$$\beta_z^I \simeq \sqrt{\varepsilon_0}k_0 + i\frac{\sqrt{\varepsilon_0}}{a\sqrt{\varepsilon_m}} \qquad (21)$$

$$\varepsilon_{eff}^{SW} \simeq \varepsilon_0 + \left[\frac{a}{d}\frac{\beta_z^I}{k_0}\tan(\beta_z^I h)\right]^2. \qquad (22)$$

Furthermore, in the case of a relatively shallow corrugation $\sqrt{\varepsilon_0}/\sqrt{\varepsilon_m} \ll \sqrt{\varepsilon_0}k_0h \ll 1$, this can be further simplified to give a relatively simple expression for the Spoof surface wave refractive index. It can be also verified that in this limit Spoof surface wave is well defined in a sense that its penetration depth into Zone II is much smaller than its propagation length

$$n_{eff}^{SW} \simeq \sqrt{\varepsilon_0} + \frac{\varepsilon_0^{3/2}}{2}\left[\frac{a}{d}k_0h\right]^2 + 2i\left[\frac{a}{d}k_0h\right]^2 \frac{\varepsilon_0^{3/2}}{k_0a\sqrt{\varepsilon_m}} \quad (23)$$

*C. Spatial localization and propagation distance of Spoof surface waves*

In Fig. 3 we study in more details the choice of design parameters and their effect on the propagation distance and localization of Spoof surface waves in the semi-infinite cladding region (Zones II). Our main question is whether Spoof surface waves are well defined. We consider that a surface wave is well defined if its propagation distance is much larger than the wave extent into a lower permittivity cladding $L_z^{II} < L_x$ as defined in (9), (10). In this study we resort to numerical solution of a full formulation (2), (3) for dispersion relation of the fundamental Spoof surface wave. Particularly, from analysis of (17-20) derived under assumption of subwavelength corrugation ($\beta_x^{SW}d \ll 1$) it follows that a critical design parameter defining Spoof surface wave loss is $1/ak_0\sqrt{\varepsilon_m}$ as it also defines radiation losses of a single groove waveguide. Next, without the loss of generality we fix $d = 1.5a$ and $\nu = 1$ THz, and assume that the filler dielectric is air $\varepsilon_0 = 1$. We then study dependence of the normalized Spoof surface propagation distance $L_x k_0/\sqrt{\varepsilon_0}$, as well as the wave penetration into the filler material $L_z^{II} k_0/\sqrt{\varepsilon_0}$ as a universal function of the corrugation depth parameter $hk_0\sqrt{\varepsilon_0}$ and the groove waveguide loss parameter $1/ak_0\sqrt{\varepsilon_m}$ (see Fig. 3). Note that in the limit of low losses $1/ak_0\sqrt{\varepsilon_m} \ll 1$ we expect the maximal confinement near the upper cut-off frequency defined by $hk_0\sqrt{\varepsilon_0} = \frac{\pi}{2}$. We also present results for several values of the high permittivity dielectric $\varepsilon_m = 100, 1000,$ and $10000$. Additionally, in Fig. 3 in gray we identify the regions where subwavelength corrugation condition $\beta_x^{SW}d < 1$ does not hold, and, therefore, the found dependencies are not universal (or might have limited validity) when using the nondimensional parameters $1/ak_0\sqrt{\varepsilon_m}$, $hk_0\sqrt{\varepsilon_0}$ (for fixed values of $a/d$ and $\varepsilon_0/\varepsilon_m$) as predicted by (19), (20). Black regions in Fig. 3 define parameter space where either numerical solution is not found or physical conditions (6), (7) for the existence of a forward propagating Spoof surface wave are not respected. Finally, dashed curves in Fig. 3 border the regions where the modal confinement in the filler dielectric becomes subwavelength $L_z^{II} = \lambda/\sqrt{\varepsilon_0}$ or where the Spoof surface wave becomes well defined $L_z^{II}/L_x = 1$. Thus, a design parameter space that would, for example, result in a well-defined Spoof surface wave with subwavelength extent into the filler dielectric will be between the two dashed lines shown in Fig. 3 (see second and third row of panels).

From the presented data we conclude that subwavelength confinement of a Spoof surface wave can be achieved near the upper cut-off frequency (larger values of $hk_0\sqrt{\varepsilon_0}$ bordering black regions in Fig. 3), when minimizing the radiation losses of a groove waveguide (smaller values of $1/ak_0\sqrt{\varepsilon_m}$). At the same time, in this region of design space propagation distance $L_x$ can become smaller than the modal transverse size $L_z^{II}$, thus rendering Spoof waves ill-defined. However, by somewhat reducing the groove depth $hk_0\sqrt{\varepsilon_0}$ (while still operating outside of the gray region) allows the propagation distance $L_x$ to increase while retaining strong modal confinement near the interface. Finally, increasing the permittivity $\varepsilon_m$, allows to reduce the propagation loss of a Spoof surface mode, extend its propagation distance $L_x$, as well as to increase the validity region of the subwavelength corrugation condition $\beta_x^{SW}d < 1$ (resulting in smaller gray areas in Fig. 3). Validity of the subwavelength corrugation condition can be also extended by choosing a deeply subwavelength period $d$.

## III. COMPARISON OF ANALYTIC AND NUMERICAL RESULTS USING IBC

In this section, we study in more detail dispersion relation of Spoof surface waves on corrugated surfaces as a function of its various parameters. Studied geometry is presented in Fig. 1 (a), with all the boundaries between the filler dielectric and the high dielectric constant material being of the IBC type. The results of semi-analytical calculations (2), (3) are verified using finite element COMSOL mode solver using Floquet boundary conditions along the modal propagation direction (see more details in Supplementary Materials). Studied geometry is The Fig. 2 compares dispersion relations of the fundamental Spoof surface wave supported by the corrugated surface made of materials with purely real and positive permittivity as obtained by solving equations (2), (3) using the Newton-Raphson iteration method (solid), with that computed using COMSOL solver (dashed). Here, we use $d = 50$ μm, $a = 35$ μm and $h = 75$ μm, and frequencies in the (0.1-1 THz) range. For the materials, we use $\varepsilon_0 = 1$ and $\varepsilon_m = 100$, which corresponds respectively to the relative dielectric permittivity of air and TiO$_2$ [67]. The upper cut-off frequency for the fundamental Spoof surface mode is given by (13) where $l_c = 0, l = 0$, and it equals to 0.95 THz. Real and imaginary parts of the corresponding dispersion relation are shown in Fig. 4 (a), while the corresponding modal propagation distance and transverse size are shown in Fig. 4 (b). We note that near the upper cut-off frequency, modal transverse size decreases dramatically and quickly becomes subwavelength (see Fig. 4 (c)). However, when operating too close to the upper cut-off frequency, there is a significant increase in the radiation loss into zone III, thus leading to much shorter propagation distances. As seen from Fig. 4, there exists an optimal spectral region where transverse modal confinement is subwavelength, while propagation distance is larger than the wave transverse size.

Next, we study the effects of the grove width to period ratio $a/d$, and groove height $h$ on the Spoof surface wave dispersion, as well as validity of the semi-analytical formulation (2), (3) when describing such waves. In Fig. 5 we present modal dispersion relation when varying groove width $a$ for a fixed period $d = 50$μm and groove height $h = 75$ μm. From these results, we observe that increasing groove width $a$ somewhat increases the upper cut-off frequency of the fundamental Spoof surface wave, while it significantly increases the modal losses. For high enough values of $a$, dispersion relation of a Spoof surface wave crosses the light line of the filler dielectric and extends further toward higher values of the propagation constant, which leads to stronger modal confinement at the interface. One way to understand this phenomenon is to note that larger groove sizes lead to stronger modal presence in the grooves, as well as stronger radiation losses at the bottom of the planar waveguides (Zone I) that radiate into Zone III through their openings of size $a$.

Finally, in Fig. 6, we study the effect of the groove height on the modal dispersion relation. Here, as expected, we observe that deeper grooves result in lower cut-off frequencies, and higher propagation losses. To appreciate the effect of the groove height on the dispersion relation, in Fig. 7 we also present the data at a fixed operational frequency 1 THz, and fixed period $d = 50$ μm and groove width $a = 25$ μm, while varying the groove height $h$. From this data we see that choosing shallow grooves (small $h$) results in longer propagation distances of a Spoof surface wave, as well as its stronger presence in the filler dielectric. Higher losses for deeper grooves are easy to understand by noting that radiation in a Spoof surface wave is partially guided along the leaky waveguide (zone I), so longer waveguides naturally result in higher (almost exponentially with $h$) losses. By increasing the groove height one can, in principle, retain acceptable modal propagation distances, while also increasing modal confinement to within several wavelengths at the structured surface.

IV.  VALIDITY OF THE IMPEDANCE BOUNDARY CONDITIONS

Up to now, the impedance boundary conditions have been used to demonstrate the possibility of propagation of Spoof surface waves along corrugated structures made of non-conductive materials. Physically, IBC limits the field presence only to the low dielectric permittivity filler region, while neglecting the effects of field penetration into the high dielectric permittivity material. In reality, the electromagnetic fields localized at the corrugated interface can penetrate into structured high dielectric permittivity material and interfere within the notches. Despite the fact that electric field will be vanishingly small compared to the magnetic field in the limit $\varepsilon_m \to +\infty$, this alone does not prevent coupling of the fields in adjacent grooves via non-zero magnetic field in the notches. Therefore, even in the limit of infinitely high $\varepsilon_m$, one still has to rely on one of the several decoherence mechanisms such as scattering, absorption or radiation (see Fig. 1) within the notches, so that the formulation using IBC remains valid. In practice, however, we often observe that numerical simulations of a complete system that includes high dielectric permittivity material indeed match those with IBC. This happens because interaction between adjacent notches is suppressed via a combination of destructive interference of the electromagnetic fields inside the notches, as well as strong radiation loss into the semi-infinite high permittivity dielectric from the narrow notches. For example, in Fig. 8 we compare dispersion relations of Spoof surface waves as computed using either a semi-analytical formulation (2), (3) with IBC, or using COMSOL with all the zones I, II, and III explicitly present in simulations (zone III is terminated with the Perfectly Matched Layer absorbing boundary domain, for more information see Supplementary Materials). As seen from the figure, at most frequencies there is a good comparison between the two dispersion relations with an exception of certain frequencies where a resonance is excited inside the notches as seeing in Fig. 8 (c). We, therefore, conclude that our semi-analytical formulation (2), (3) derived using IBC, can indeed describe dispersion relation of Spoof surface waves propagating on non-conductive lossless corrugated surfaces. This semi-analytical model, however, is limited when adjacent grooves are strongly coupled via resonant field excitation inside of the high permittivity notches.

V.  CONCLUSION

In this paper we demonstrate a new type of Spoof surface waves that propagate along periodic corrugated surfaces made of non-conductive lossless materials featuring high positive permittivity. Existence of such waves is due to possibility of the electromagnetic energy confinement in the planar groove waveguides limited by the high dielectric permittivity walls. Unlike the case of Spoof surface waves on structured metallic surfaces where electromagnetic energy is exponentially fast decreasing into metal, in our case, significant radiation presence is possible in the high dielectric permittivity material. Thus, a careful consideration of radiative effects is important for the proper design of Spoof surface waves on the non-conductive lossless surfaces. An analytic derivation of the Spoof surface wave dispersion relation is presented in the case of deeply subwavelength period of a corrugated structure using impedance boundary conditions at the interfaces. Thus, obtained dispersion relation is then verified numerically using COMSOL Multiphysics software on a full system comprising a filler dielectric, a structured high permittivity dielectric, as well as radiation absorbing boundary conditions, and an overall good agreement between semi-analytical theory and numerical simulations is obtained. Finally, we confirmed that there exists a region of design space where Spoof surface wave on a non-conducting interface can be tightly confined at such an interface with its transverse size comparable or even smaller that the wavelength of light in the filler dielectric. Additionally, such surface waves can feature propagation distances much longer than their transverse size, thus making such surface waves like the regular guided modes for practical purposes. As Spoof surface waves require high dielectric permittivity materials for their experimental realisation, we believe that longer wavelengths spectral ranges (THz, microwave, etc.) are the most suitable for the demonstration of such waves due to abundance of dielectrics with relative permittivity higher than 100.

## VI. Supplemental material

### A. Spatial localization and propagation distance of Spoof surface waves

In the grooves (zone I, in Fig. 1), we have two waves propagating along the z-direction with propagation constant denoted as $\beta_z^I$ and $-\beta_z^I$. The wave propagating toward the positive z-direction come from the reflection at the bottom of the groove at $z = -h$. Also, the structure in zone I is similar to a waveguide made of two parallel planar materials of permittivity $\varepsilon_m$ separated by a dielectric of permittivity $\varepsilon_0$. Thus, for a TM mode inside the groove waveguide, one can easily solve the Maxwell equations and find the following magnetic and electric components [68]

$$H_y^{I+} = A^+ \cos(k_x^I x)\, e^{i\beta_z^I z} \quad (24)$$
$$E_z^{I+} = -iA^+ \frac{k_x^I}{\varepsilon_0 k_0} \sin(k_x^I x)\, e^{i\beta_z^I z} \quad (25)$$
$$E_x^{I+} = A^+ \frac{\beta_z^I}{\varepsilon_0 k_0} \cos(k_x^I x)\, e^{i\beta_z^I z}, \quad (26)$$

where $A^+$ is a constant and where the relation between $k_x^I$ and $\beta_z^I$ is given by

$$k_x^I = \sqrt{\varepsilon_0 k_0^2 - (\beta_z^I)^2}. \quad (27)$$

Here for simplicity, we only consider even modes of a groove waveguide as described by (25)-(27), that also comprise the fundamental mode. While the formulation for the odd modes ($H_y^{I+} \sim \sin(k_x^I x)$) can be easily derived, in this work we do not present it for the sake of brevity. Moreover, in practical applications, one would most probably favor the fundamental even mode, as it is the easiest to be realised experimentally.

We also have three impedance boundary conditions in the groove to respect at the interfaces $x = a/2$, $x = -a/2$ and $z = -h$

$$H_y^{I+}|_{x=a/2} = -\sqrt{\varepsilon_m} E_z^{I+}|_{x=a/2} \quad (28)$$
$$H_y^{I+}|_{x=-a/2} = \sqrt{\varepsilon_m} E_z^{I+}|_{x=-a/2} \quad (29)$$
$$H_y^{I+}|_{z=-h} = -\sqrt{\varepsilon_m} E_x^{I+}|_{z=-h}. \quad (30)$$

Using the impedance boundary conditions (28) with the expressions (24) and (25), we get

$$\cos\left(k_x^I \frac{a}{2}\right) = i\sqrt{\varepsilon_m}\frac{k_x^I}{\varepsilon_0 k_0}\sin\left(k_x^I \frac{a}{2}\right) \quad (31)$$

$$\Rightarrow k_x^I \tan(k_x^I a/2) = -i\frac{k_0 \varepsilon_0}{\sqrt{\varepsilon_m}}, \quad (32)$$

where equation (32) with the relation (27) can be simplified as follows when the argument $k_x^I a/2$ is small

$$(k_x^I)^2 \simeq -i\frac{2}{a}\frac{\varepsilon_0 k_0}{\sqrt{\varepsilon_m}} \quad (33)$$

$$\Rightarrow \beta_z^I = \sqrt{\varepsilon_0 k_0^2 - (k_x^I)^2} \simeq \sqrt{\varepsilon_0} k_0 + i\frac{\sqrt{\varepsilon_0}}{a\sqrt{\varepsilon_m}}. \quad (34)$$

A numerical method such as the Newton-Raphson iteration method can also be used to find the expression of $\beta_z^I$ from equation (27) and (32). Moreover, the other wave propagating toward the negative z direction has a propagation constant that corresponds to $-\beta_z^I$. The magnetic and electric components are then given by

$$H_y^{I-} = A^- \cos(k_x^I x)\, e^{-i\beta_z^I z} \quad (35)$$
$$E_z^{I-} = -iA^- \frac{k_x^I}{\varepsilon_0 k_0}\sin(k_x^I x)\, e^{-i\beta_z^I z} \quad (36)$$
$$E_x^{I-} = -A^- \frac{\beta_z^I}{\varepsilon_0 k_0}\cos(k_x^I x)\, e^{-i\beta_z^I z}, \quad (37)$$

where $A^-$ is a constant. The total electric and magnetic components in the first zone can be written as the combination of the two waves propagating along the z-direction

$$H_y^I = \cos(k_x^I x)\left[A^+ e^{i\beta_z^I z} + A^- e^{-i\beta_z^I z}\right] \quad (38)$$
$$E_z^I = -i\frac{k_x^I}{\varepsilon_0 k_0}\sin(k_x x)\left[A^+ e^{i\beta_z^I z} + A^- e^{-i\beta_z^I z}\right] \quad (39)$$
$$E_x^I = \frac{\beta_z^I}{\varepsilon_0 k_0}\cos(k_x^I x)\left[A^+ e^{i\beta_z^I z} - A^- e^{-i\beta_z^I z}\right]. \quad (40)$$

In addition, the relation between the coefficients $A^+$ and $A^-$ can be found with the impedance boundary condition at $z = -h$ and equations (38) and (40):

$$A^+ e^{-i\beta_z^I h}\left(\frac{\beta_z^I}{\varepsilon_0 k_0} + \frac{1}{\sqrt{\varepsilon_m}}\right) = A^- e^{i\beta_z^I h}\left(\frac{\beta_z^I}{\varepsilon_0 k_0} - \frac{1}{\sqrt{\varepsilon_m}}\right) \quad (41)$$

$$\Rightarrow A^+ = A^- e^{2i\beta_z^I h}\left(\frac{\sqrt{\varepsilon_m}\beta_z^I - \varepsilon_0 k_0}{\sqrt{\varepsilon_m}\beta_z^I + \varepsilon_0 k_0}\right). \quad (42)$$

Because of the impedance boundary condition and radiative loss at the waveguide boundaries, the amplitude of the wave propagating toward the positive z direction is smaller than the wave propagating toward the bottom of the cavity. Also, the groove depth $h$ controls the phase difference between these two waves. These observations suggest that the groove depth will have a significant impact on the complex dispersion relation of the Spoof surface wave supported by the corrugated surface.

Next, the magnetic and electric components of the wave propagating along the x axis in the region above the surface ($z > 0$) are given by:

$$H_y^{II} = \sum_n B_n e^{i\beta_x^{II,n} x}\, e^{ik_z^{II,n} z} \quad (43)$$
$$E_x^{II} = \sum_n B_n \frac{k_z^{II,n}}{\varepsilon_0 k_0} e^{i\beta_x^{II,n} x}\, e^{ik_z^{II,n} z} \quad (44)$$
$$E_z^{II} = -\sum_n B_n \frac{\beta_x^{II,n}}{\varepsilon_0 k_0} e^{i\beta_x^{II,n} x}\, e^{ik_z^{II,n} z} \quad (45)$$
$$(k_z^{II,n})^2 = \varepsilon_0 k_0^2 - (\beta_x^{II,n})^2;\ \beta_x^{II,n} = \beta_x^{SW} + G_n, \quad (46)$$

where $G_n = 2\pi n/d$ comes from the Bloch periodicity. At $z = 0$, we must match the components from zone I and II above the grooves $x \in (-a/2, a/2)$ and respect the impedance boundary condition otherwise for $x \in (-d/2, -a/2) \cap (a/2, d/2)$. For the first case above the grooves, we have

$$H_y^{II}|_{z=0} = H_y^I|_{z=0} \quad (47)$$
$$E_x^{II}|_{z=0} = E_x^I|_{z=0}. \quad (48)$$

Then, we have for the impedance boundary condition at the interface at $x \in (-d/2, -a/2) \cap (a/2, d/2)$

$$H_y^{II}|_{z=0} = -\sqrt{\varepsilon_m} E_x^{II}|_{z=0} \quad (49)$$

Here, to find the expression of the effective refractive index of the Spoof surface waves, one must solve the previous equations and conditions. To begin, a first relation between the coefficients $B_n$, $A^+$ and $A^-$ can be found from relations (48) and (49)

$$E_x^{II}|_{z=0} = \sum_n B_n \frac{k_z^{II,n}}{\varepsilon_0 k_0} e^{i\beta_x^{II,n} x} =$$
$$\begin{cases} A \frac{\beta_z^I}{\varepsilon_0 k_0} \cos(k_x^I x), & x \in \left(-\frac{a}{2}, \frac{a}{2}\right) \\ -\frac{1}{\sqrt{\varepsilon_m}} \sum_n B_n e^{i\beta_x^{II,n} x}, & x \in \left(-\frac{d}{2}, -\frac{a}{2}\right) \cap \left(\frac{a}{2}, \frac{d}{2}\right), \end{cases} \quad (50)$$

where $A = A^+ - A^-$. In the limit of $d \ll \lambda$, only $n = 0$ is of importance. Thus, we get

$$E_x^{II}|_{z=0} = B_0 k_z^{SW,II} e^{i\beta_x^{SW} x} =$$
$$\begin{cases} A\beta^I \cos(k_x^I x), & x \in \left(-\frac{a}{2}, \frac{a}{2}\right) \\ -\frac{\varepsilon_0 k_0}{\sqrt{\varepsilon_m}} B_0 e^{i\beta_x^{SW} x}, & x \in \left(-\frac{d}{2}, -\frac{a}{2}\right) \cap \left(\frac{a}{2}, \frac{d}{2}\right) \end{cases} \quad (51)$$

$$\Rightarrow k_z^{SW,II} \int_{-d/2}^{d/2} e^{i\beta_x^{SW} x} dx = \beta_z^I \frac{A}{B_0} \int_{-a/2}^{a/2} \cos(k_x^I x) dx - \frac{\varepsilon_0 k_0}{\sqrt{\varepsilon_m}} \left[ \int_{-\frac{d}{2}}^{-\frac{a}{2}} e^{i\beta_x^{SW} x} dx + \int_{\frac{a}{2}}^{\frac{d}{2}} e^{i\beta_x^{SW} x} dx \right] \quad (52)$$

$$\Rightarrow k_z^{SW,II} \sin\left(\beta_x^{SW} \frac{d}{2}\right) = \frac{A}{B_0} \frac{\beta_z^I \beta_x^{SW}}{k_x^I} \sin\left(k_x^I \frac{a}{2}\right) - \frac{\varepsilon_0 k_0}{\sqrt{\varepsilon_m}} \left[ \sin\left(\beta_x^{SW} \frac{d}{2}\right) - \sin\left(\beta_x^{SW} \frac{a}{2}\right) \right] \quad (53)$$

$$\Rightarrow B_0 = A \frac{\beta_z^I \beta_x^{SW}}{k_x^I} \frac{\sin\left(k_x^I \frac{a}{2}\right)}{\left(\frac{\varepsilon_0 k_0}{\sqrt{\varepsilon_m}} + k_z^{SW,II}\right) \sin\left(\beta_x^{SW,II} \frac{d}{2}\right) - \frac{\varepsilon_0 k_0}{\sqrt{\varepsilon_m}} \sin\left(\beta_x^{SW} \frac{a}{2}\right)}. \quad (54)$$

A third relation between $A^+$, $A^-$ and $B_0$ can be found with the continuity condition of the magnetic component at $z = 0$ represented by equation (47)

$$\sum_n B_n e^{i\beta_x^{II,n} x} = (A^+ + A^-) \cos(k_x^I x). \quad (55)$$

In the limit of $d \ll \lambda$, only $n = 0$ is of importance. So, we get:

$$B_0 e^{i\beta_x^{SW} x} = (A^+ + A^-) \cos(k_x^I x) \quad (56)$$

$$\Rightarrow B_0 \int_{-a/2}^{a/2} e^{i\beta_x^{SW} x} dx = (A^+ + A^-) \int_{-a/2}^{a/2} \cos(k_x^I x) dx \quad (57)$$

$$\Rightarrow B_0 \frac{1}{\beta_x^{SW}} \sin\left(\beta_x^{SW} \frac{a}{2}\right) = (A^+ + A^-) \frac{1}{k_x^I} \sin\left(k_x^I \frac{a}{2}\right) \quad (58)$$

$$\Rightarrow B_0 = (A^+ + A^-) \frac{\beta_x^{SW}}{k_x^I} \frac{\sin\left(k_x^I \frac{a}{2}\right)}{\sin\left(\beta_x^{SW} \frac{a}{2}\right)}. \quad (59)$$

At this point we have 3 unknow constants $A^+$, $A^-$ and $B_0$ with three equations (42), (54) and (59). Thus, $\beta_x^{SW}$ can now be solved

$$(A^+ + A^-) \frac{\beta_x^{SW}}{k_x^I} \frac{\sin\left(k_x^I \frac{a}{2}\right)}{\sin\left(\beta_x^{SW} \frac{a}{2}\right)} =$$
$$(A^+ - A^-) \frac{\beta_z^I \beta_x^{SW}}{k_x^I} \frac{\sin\left(k_x^I \frac{a}{2}\right)}{\left(\frac{\varepsilon_0 k_0}{\sqrt{\varepsilon_m}} + k_z^{SW,II}\right) \sin\left(\beta_x^{SW} \frac{d}{2}\right) - \frac{\varepsilon_0 k_0}{\sqrt{\varepsilon_m}} \sin\left(\beta_x^{SW} \frac{a}{2}\right)} \quad (60)$$

$$\Rightarrow \left(\frac{\varepsilon_0 k_0}{\sqrt{\varepsilon_m}} + k_z^{SW,II}\right) \frac{\sin\left(\beta_x^{SW} \frac{d}{2}\right)}{\sin\left(\beta_x^{SW} \frac{a}{2}\right)} =$$
$$\beta_z^I \frac{i\sqrt{\varepsilon_m} \beta_z^I \tan(\beta_z^I h) - \varepsilon_0 k_0}{\sqrt{\varepsilon_m} \beta_z^I - i\varepsilon_0 k_0 \tan(\beta_z^I h)} + \frac{\varepsilon_0 k_0}{\sqrt{\varepsilon_m}}, \quad (61)$$

where $k_z^{SW,II} = \sqrt{\varepsilon_0 k_0^2 - (\beta_x^{SW})^2}$ and $\beta_x^{SW} = k_0 n_{eff}^{SW}$. Up to now, we found analytic expressions of the dispersion relations of the wave propagating in the grooves (32) and the Spoof surface waves propagating along the interface (61). One can then use numerical method such as the Newton-Raphson iteration method to solve for $\beta_z^I$ and $\beta_x^{SW}$. However, further simplifications can be applied under specific conditions. To begin with, the dispersion relation can be greatly simplified while considering an ideal case where the corrugated structure is made of material with very high positive permittivity $\varepsilon_m \to +\infty$

$$\beta_x^{SW} = \sqrt{\varepsilon_0 k_0^2 + \left(\beta_z^I \tan(\beta_z^I h) \frac{\sin\left(\beta_x^{SW} \frac{a}{2}\right)}{\sin\left(\beta_x^{SW} \frac{d}{2}\right)}\right)^2}. \quad (62)$$

Additionally, further simplifications can be done for a subwavelength corrugation $\beta_x^{SW} d \ll 1$

$$\beta_x^{SW} = \sqrt{\varepsilon_0 k_0^2 + \left(\beta_z^I \tan(\beta_z^I h) \left(\frac{a}{d}\right)\right)^2}. \quad (63)$$

Both equations can be represented as universal functions of several non-dimensional variables, which will help to identify the relation of each parameter of the structured material $d$, $a$, $h$, $\sqrt{\varepsilon_0}$, and $\sqrt{\varepsilon_m}$ on the dispersion relations. In fact, we can show that we can represent the effective refractive index of the waves $n_{eff}^I$ and $n_{eff}^{SW}$ as a function of these following non-dimensional variables $hk_0\sqrt{\varepsilon_0}$, $ak_0\sqrt{\varepsilon_0}$, $dk_0\sqrt{\varepsilon_0}$, and $\sqrt{\varepsilon_0}/\sqrt{\varepsilon_m}$

$$\frac{k_x^I}{\sqrt{\varepsilon_0} k_0} \tan\left(\frac{k_x^I}{\sqrt{\varepsilon_0} k_0} \frac{ak_0\sqrt{\varepsilon_0}}{2}\right) = -i \frac{\sqrt{\varepsilon_0}}{\sqrt{\varepsilon_m}} \quad (64)$$

$$\Rightarrow \frac{n_{eff}^I}{\sqrt{\varepsilon_o}} = \pm \sqrt{1 - \left(\frac{k_x^I}{\sqrt{\varepsilon_o}k_0}\right)^2} \quad (65)$$

$$\left(\frac{\sqrt{\varepsilon_o}}{\sqrt{\varepsilon_m}} + \frac{k_z^{SW,II}}{\sqrt{\varepsilon_o}k_0}\right) \frac{\sin\left(\frac{n_{eff}^{SW}}{\sqrt{\varepsilon_o}}\cdot\frac{dk_0\sqrt{\varepsilon_o}}{2}\right)}{\sin\left(\frac{n_{eff}^{SW}}{\sqrt{\varepsilon_o}}\cdot\frac{ak_0\sqrt{\varepsilon_o}}{2}\right)} =$$

$$i\frac{\beta_z^I}{\sqrt{\varepsilon_o}k_0}\frac{\tan\left(\frac{n_{eff}^I}{\sqrt{\varepsilon_o}}hk_0\sqrt{\varepsilon_o}\right)+i\frac{\sqrt{\varepsilon_o}}{\sqrt{\varepsilon_m}}\cdot\frac{\sqrt{\varepsilon_o}}{n_{eff}^I}}{1-i\frac{\sqrt{\varepsilon_o}}{\sqrt{\varepsilon_m}}\cdot\frac{\sqrt{\varepsilon_o}}{n_{eff}^I}\tan\left(\frac{n_{eff}^I}{\sqrt{\varepsilon_o}}hk_0\sqrt{\varepsilon_o}\right)} + \frac{\sqrt{\varepsilon_o}}{\sqrt{\varepsilon_m}} \quad (66)$$

$$\Rightarrow \frac{k_z^{SW,II}}{\sqrt{\varepsilon_o}k_0} = \pm\sqrt{1-\left(\frac{n_{eff}^{SW}}{\sqrt{\varepsilon_o}}\right)^2}. \quad (67)$$

*B. Details of COMSOL simulations*

To validate the analytic dispersion relation given by equations (2)-(3) along with the study of the validity of the IBC, the finite element software COMSOL Multiphysics 5.5 has been used to numerically solve the electromagnetic surface waves supported by the corrugated structure. All the simulations were performed in the frequency domain, where one period of the structure is modeled together with the Floquet periodic conditions at $x = \pm d/2$. This condition along with the COMSOL's eigenfrequency study step allows us to find the complex eigen frequency $v_{num}(\beta_x)$ that corresponds to a given real propagation constant $\beta_x$. For the structure using IBC, a simple domain made of the filler dielectric of dielectric permittivity $\varepsilon_0$ representing the combination of region I and II is delimited by the periodic conditions (region I) and IBC inside the groove (region II). The region above $z = 0$, which extends up to multiple times the operational wavelengths, is terminated with a PML (perfect matched layers) domain and scattering boundary condition at the outer boundary. For the full structure (without IBC), all the three zones illustrated in Fig. 1 along with the high dielectric permittivity material $\varepsilon_m$ in zone III and below $z = -h$ is included into simulation cell. In this case, the computational cell is terminated both from the top and bottom with the PML domains and scattering boundary conditions.

As mentioned earlier, the eigenfrequency search method was used to numerically find the complex dispersion relation $v_{num}(\beta_x) = v_{num}^r(\beta_x) + iv_{num}^i(\beta_x)$ of the supported surface waves, where $\beta_x$ is real and specified in the Floquet boundary conditions. In practical applications, however, one is rather interested in the complex propagation constant of the mode $\beta_x^{SW}(v) = \beta_x^{SW,r}(v) + i\beta_x^{SW,i}(v)$ for the pure real frequencies $v$. Thus, for practical applications, and in order to compare semi-analytic dispersion relation given by (2)-(3) along with the numerical ones, one has to know how to transform dispersion relation in terms of a complex frequency $v_{num}(\beta_x)$ given by COMSOL into dispersion relations in terms of a complex propagation constant $\beta_x^{SW}(v)$. Particularly, assuming that eigen frequency $v_{num}(\beta_x)$ (as computed by COMSOL) can be computed for any propagation constant $\beta_x$, we then look for the complex propagation constants $\beta_x$ that result in the purely real eigenfrequencies

$$v_{num}^i(\beta_x^r, \beta_x^i) = 0 \quad (68)$$

Multiple methods can be used to solve (68). As example, one could start with a purely real propagation constant $\beta_{x,0} = (\beta_x^r, 0)$ that would result in the complex eigenfrequency $v_{num}(\beta_{x,0})$, and then use an iterative method (like the method of secants used in our work) to find the corresponding imaginary part of the propagation constant $\beta_x^i$ so that the complex $\beta_x = (\beta_x^r, \beta_x^i)$ would result in the purely real eigen frequency (68). Particularly, for the $n^{th}$ iteration one can use

$$\beta_{x,n+1}^i = \beta_{x,n}^i - v_{num}^i(\beta_x^r, \beta_{x,n}^i)\frac{\beta_{x,n}^i - \beta_{x,n-1}^i}{v_{num}^i(\beta_x^r, \beta_{x,n}^i) - v_{num}^i(\beta_x^r, \beta_{x,n-1}^i)} \quad (69)$$

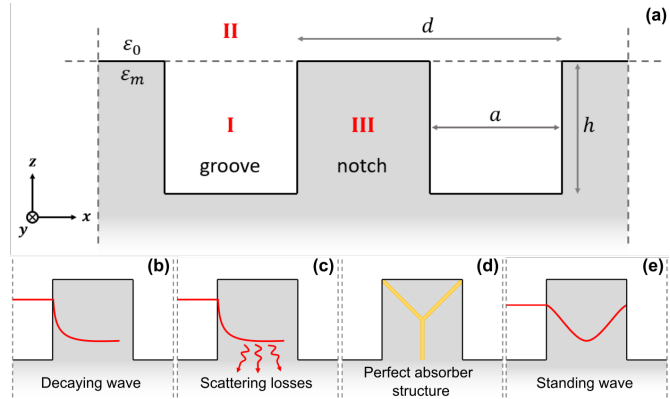

Fig. 2. Schematic of the corrugated interface, and validity of the Impedance Boundary Condition. (a) Schematic of the corrugated interface. The period is d and the width and height of the grooves are respectively a and h. The permittivity of the dielectric filler above the corrugated surface (zones II) and in the grooves (zone I) is $\varepsilon_0$, while permittivity of the corrugated surface (zone III) is $\varepsilon_m$. (b)-(e) Validity of the Impedance Boundary Condition demands that the outgoing wave from the interface of the grooves (into notches) is either absorbed (b) or scattered (b) or suppressed (c), (d) from interacting with other interfaces within the notches.

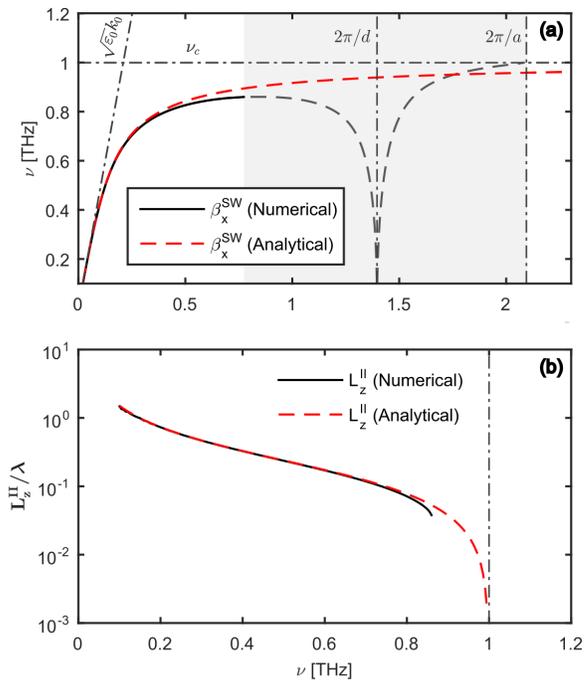

Fig. 1. Dispersion relation of a Spoof surface wave for the case of Perfect Isolator $\varepsilon_m = +\infty$. (a) Dispersion relation of a Spoof surface wave using analytical approximation (14) (dashed red) and by numerically solving equation (12) (solid black) with the Newton-Raphson iteration method. (b) Penetration depth of the Spoof surface wave into filler dielectric (Zone II) as defined by the analytical expression (15) (dashed red) and from numerical solution (12) (solid black) in the units of a free space wavelength $\lambda$. Here, $d = 45$ μm, $a = 30$ μm and $h = 75$ μm.

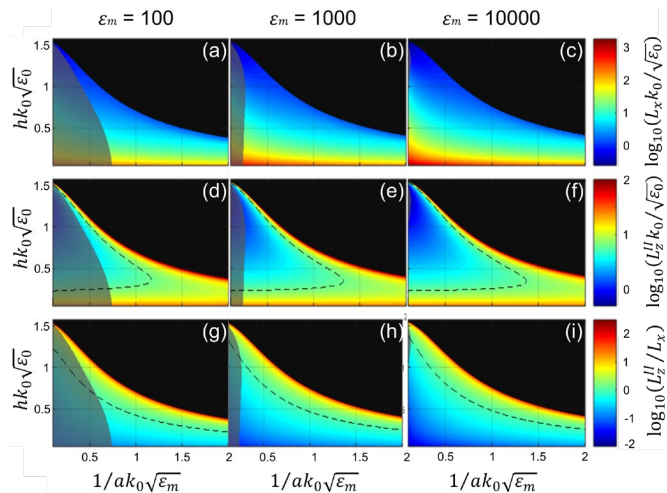

Fig. 4. Various properties and regimes of Spoof surface wave supported by the corrugated structure. (a)-(c) Normalized propagation distance $L_x k_0/\sqrt{\varepsilon_0}$, (d)-(f) Normalized penetration into the filler dielectric material $L_z^{II} k_0/\sqrt{\varepsilon_0}$, (g)-(i) the ratio between the penetration into the filler dielectric material $L_z^{II}$ and the propagation distance $L_x$ are presented in function of the parameters $1/ak_0\sqrt{\varepsilon_m}$ and $hk_0\sqrt{\varepsilon_0}$ for respectively $\varepsilon_m = 100$, $\varepsilon_m = 1000$, and $\varepsilon_m = 10000$. The regions in gray indicate non-subwavelength regime. The dashed lines indicate (d)-(f) the subwavelength limit $L_z^{II} k_0\sqrt{\varepsilon_0} = 2\pi$ and (g)-(i) the well-defined surface wave limit $L_z^{II}/L_x = 1$. Here, the operational frequency is 1 THz and $d = 1.5a$.

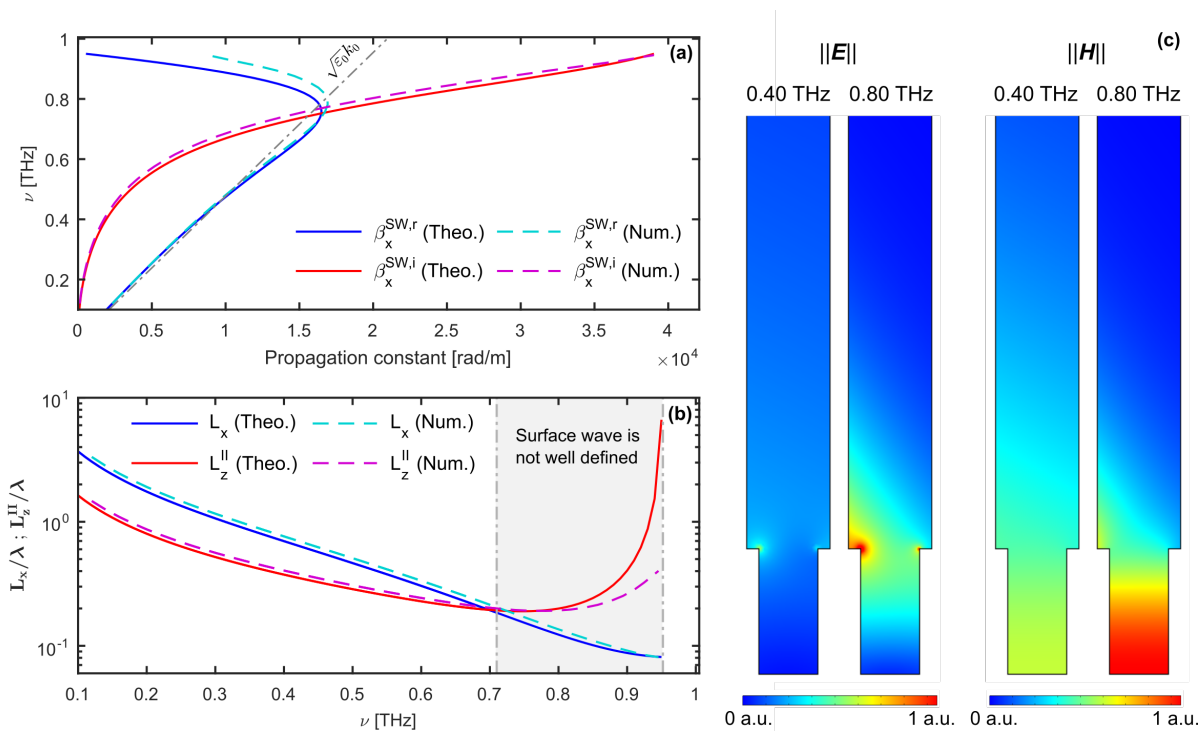

Fig. 3. Typical dispersion relation of a Spoof surface wave supported by the corrugated surface. (a) Complex dispersion relation of a fundamental Spoof surface wave as found by solving (2), (3) using Newton-Raphson iteration method (solid curves) are compared to those found by COMSOL (dashed curves). (b) Propagation distance $L_x$ (blue, cyan) and penetration depth into air $L_z^{II}$ (red, magenta) in units of λ. (c) Distribution of the electric and magnetic field amplitudes in air at 0.40 THz and 0.80 THz found using COMSOL. Here, $d = 50$ μm, $a = 35$ μm, $h = 75$ μm, $\varepsilon_0 = 1$ and $\varepsilon_m = 100$.

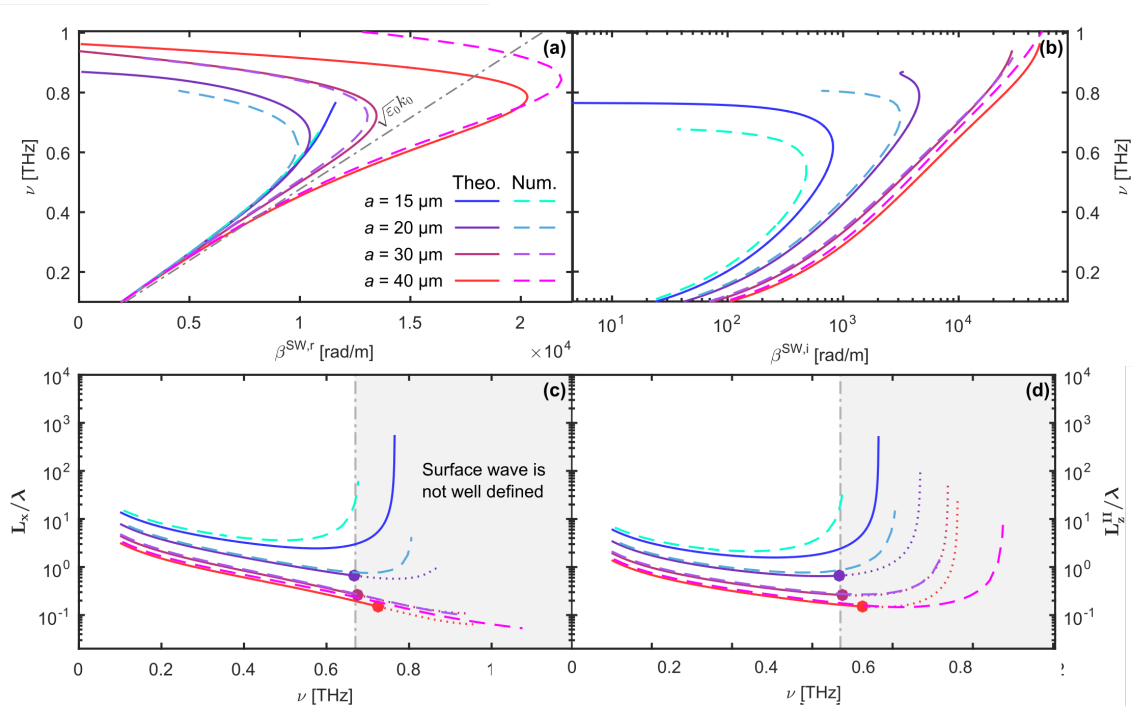

Fig. 5. Propagation constant of a Spoof surface wave for various groove width $a$. (a) Real part and (b) Imaginary part of the propagation constant of the fundamental Spoof surface wave for three structures of increasing groove width $a = 15$ μm, $a = 20$ μm, $a = 30$ μm, and $a = 40$ μm. (c) Propagation distance $L_x$ and (d) penetration depth into air $L_z^{II}$ in units of $\lambda$. Results are found by solving (2), (3) using Newton-Raphson iteration method (solid curves) are compared to those found by COMSOL (dashed curves). The circles presented in (c) and (d) indicate the limit where the surface wave is well-defined $L_x/L_z^{II} = 1$. The period $d$ is 50 μm, the groove depth $h$ is 75 μm and the permittivity constants are $\varepsilon_0 = 1$ and $\varepsilon_m = 100$.

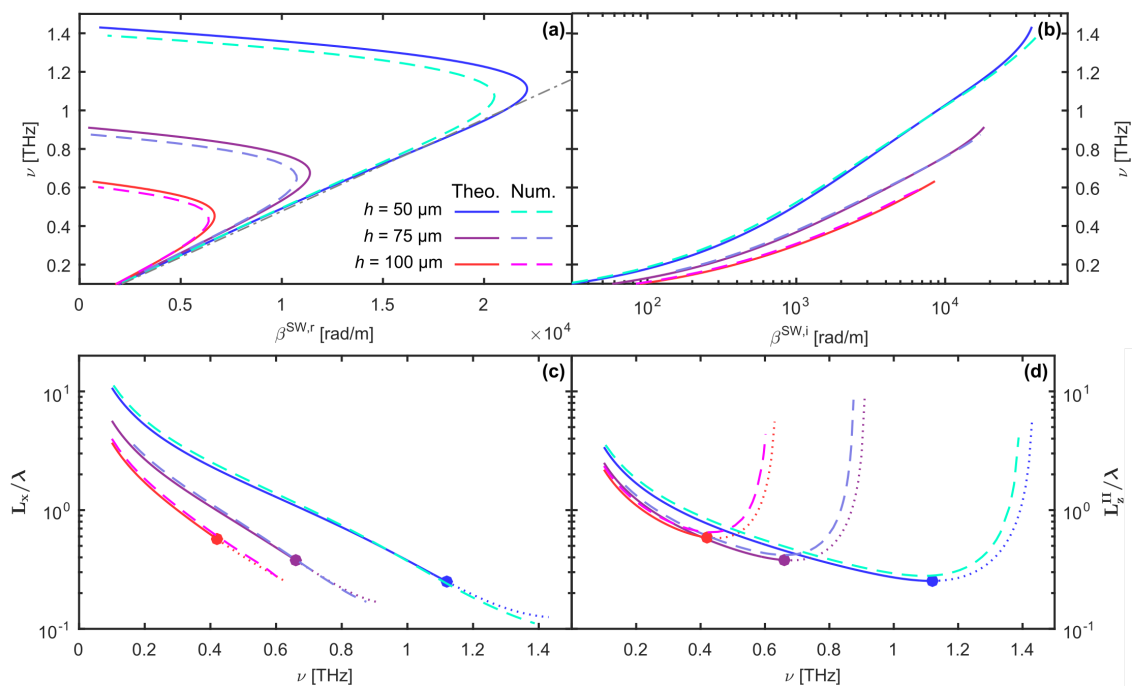

Fig 6. Propagation constant of a Spoof surface wave for various groove height $h$. (a) Real part and (b) Imaginary part of the propagation constant of the fundamental Spoof surface wave for three structures of increasing groove height $h = 50$ μm, $h = 75$ μm and $h = 100$ μm. (c) Propagation distance $L_x$ and (d) penetration depth into air $L_z^{II}$ in units of $\lambda$. Results are found by solving (2), (3) using Newton-Raphson iteration method (solid curves) are compared to those found by COMSOL (dashed curves). The circles presented in (c) and (d) indicate the limit where the surface wave is well-defined $L_x/L_z^{II} = 1$. The period $d$ is 50 μm, the groove width $a$ is 25 μm and the permittivity constants are $\varepsilon_0 = 1$ and $\varepsilon_m = 100$.

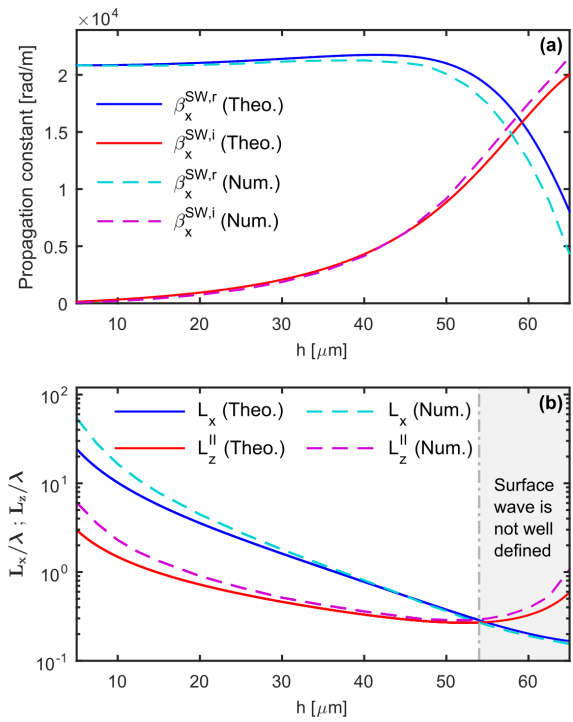

Fig 7. Propagation constant and modal sizes as a function of the groove depth $h$ at 1THz operation frequency. (a) Real and imaginary parts of the fundamental Spoof surface wave propagation constant as a function of the groove depth. (b) Propagation distance $L_x$ and penetration depth into air $L_z^{II}$ in units of $\lambda$. Here, $a = d/2$, $d = 50\ \mu m$ and the permittivity constants are $\varepsilon_0 = 1$ and $\varepsilon_m = 100$.

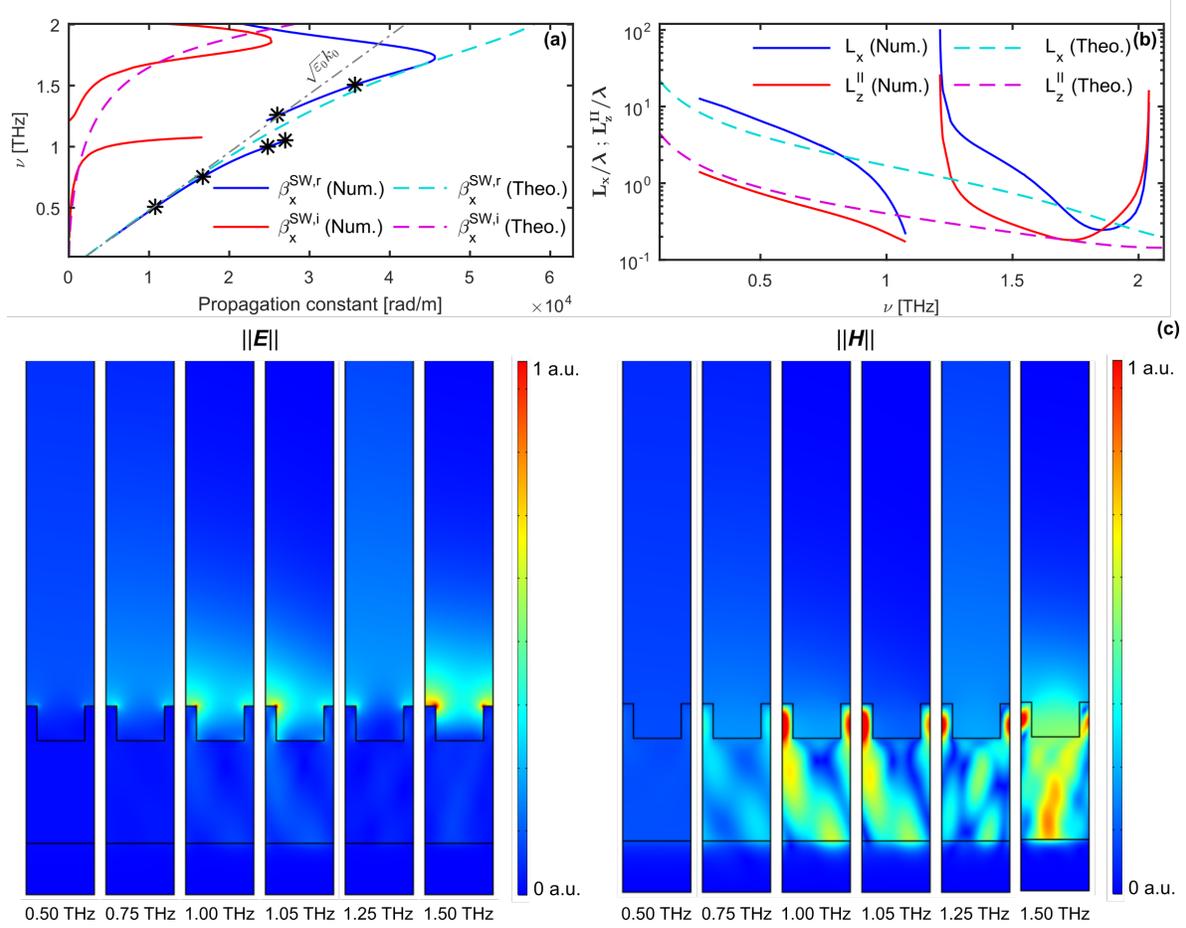

Fig. 8. Dispersion relation of Spoof surface waves computed using full geometry of the corrugated surface (zones I, II, and III, no IBC) terminated with a Perfectly Matched Layer. (a) Real (blue, cyan) and imaginary (reg, magenta) parts of the Spoof surface wave propagation constant as found using COMSOL simulation of a complete geometry (solid), and as computed equations (2), (3) (dotted). (b) Propagation distance (blue, cyan) and penetration depth into air (red, magenta) in units of $\lambda$. (c) Distribution of the electric and magnetic fields at various frequencies 0.50 THz (no resonance in the notch), 0.75 THz, 1.00 THz (resonance in the notch), 1.05 THz, 1.25 THz, and 1.50 THz. Here, $d = 50$ μm, $a = 35$ μm and $h = 25$ μm, $\varepsilon_0 = 1$ and $\varepsilon_m = 100$.